\documentclass[letterpaper]{article} 
\usepackage{aaai2026}  
\usepackage{times}  
\usepackage{helvet}  
\usepackage{courier}  
\usepackage[hyphens]{url}  
\usepackage{graphicx} 
\usepackage{natbib}  
\usepackage{caption} 
\usepackage{algorithm}
\usepackage{algorithmic}

\usepackage{booktabs}
\usepackage{array}
\usepackage{siunitx}
\usepackage{caption}
\usepackage{multirow}
\usepackage{makecell}
\usepackage{xcolor}
\usepackage{colortbl}
\usepackage{booktabs}
\usepackage{float}
\usepackage{amssymb}
\usepackage{newfloat}
\usepackage{listings}
\DeclareCaptionStyle{ruled}{labelfont=normalfont,labelsep=colon,strut=off} 
\floatstyle{ruled}
\newfloat{listing}{tb}{lst}{}
\floatname{listing}{Listing}
 \nocopyright 

\title{Is-NeRF: In-scattering Neural Radiance Field for Blurred Images}
\author{
    Nan Luo\textsuperscript{\rm 1}\thanks{Corresponding author}, Chenglin Ye\textsuperscript{\rm 2}, Jiaxu Li\textsuperscript{\rm 1}, Gang Liu\textsuperscript{\rm 1}, Bo Wan\textsuperscript{\rm 1}, Di Wang\textsuperscript{\rm 1}, Lupeng Liu\textsuperscript{\rm 2}, Jun Xiao\textsuperscript{\rm 2}
}
\affiliations{
    \textsuperscript{\rm 1}Xidian University\\
    \textsuperscript{\rm 2}University of the Chinese Academy of Sciences\\

}

\usepackage{bibentry}

\begin{document}

\maketitle

\begin{abstract}
Neural Radiance Fields (NeRF) has gained significant attention for its prominent implicit 3D representation and realistic novel view synthesis capabilities. Available works unexceptionally employ straight-line volume rendering, which struggles to handle sophisticated lightpath scenarios and introduces geometric ambiguities during training, particularly evident when processing motion-blurred images. To address these challenges, this work proposes a novel deblur neural radiance field, Is-NeRF, featuring explicit lightpath modeling in real-world environments. By unifying six common light propagation phenomena through an in-scattering representation, we establish a new scattering-aware volume rendering pipeline adaptable to complex lightpaths. Additionally, we introduce an adaptive learning strategy that enables autonomous determining of scattering directions and sampling intervals to capture finer object details. The proposed network jointly optimizes NeRF parameters, scattering parameters, and camera motions to recover fine-grained scene representations from blurry images. Comprehensive evaluations demonstrate that it effectively handles complex real-world scenarios, outperforming state-of-the-art approaches in generating high-fidelity images with accurate geometric details. 
\end{abstract}

\section{Introduction}
The emergence of Neural Radiance Field \cite{mildenhall2021nerf}  (NeRF) has catalyzed extensive research efforts, driven by its remarkable capability of implicit 3D representation and realistic novel view synthesis.
NeRF maps arbitrary continuous 3D coordinates to volume densities and colors via MLPs given a set of posed images, and then generates novel view images via volume rendering. While NeRF has achieved great success in various domains \cite{barron2021mip, martin2021nerf, huang2022hdr, levy2023seathru}, most variants are trained and tested under meticulously controlled environments with images clearly captured from multiple viewpoints. In real-world scenarios, image data usually contains various forms of degradation or noise, such as motion blur or defocus that degrade image quality, thereby diminishing NeRF's performance. 

Recent studies have explored extending NeRF to address motion-blurred image reconstruction \cite{ma2022deblur, lee2023dp, wang2023bad, 10.1145/3664647.3680569, 2025Deblur}, achieving partial improvements in implicit 3D scene representation. However, the presence of non-Lambertian surfaces in real-world scenes (e.g. glasses, metallic materials, and liquid interfaces) brings difficulties to the standard volume rendering paradigm adopted by NeRF-based methods. Due to the light reflection or refraction, the color of spatial point at the surface of medium varies sharply with respect to viewing directions, which corresponds to high frequency property in color variation that can not be precisely learned and expressed by neural networks, resulting in ambiguities in color and geometry during training. Such deficiencies become particularly pronounced under motion blur conditions, where conventional positional encoding techniques prove insufficient to resolve inherent variations in blurred observations. Furthermore, the rigid sampling strategy of NeRF, despite coarse-to-fine design, lacks the adaptability required to preserve geometric details in small-scale objects or thin structures. 

In tackling these limitations, this paper attempts to model complex light transport in real-world scenes based on light scattering theory. By simplifying the commonly encountered light propagation phenomena (including reflection, refraction, transmission, glossiness, diffusion) with a generalized in-scattering model, we propose a new scattering-aware volume rendering method and an optimized neural radiance field network for motion blur mitigation. Our network simultaneously learns NeRF parameters, scattering parameters, and camera motions, achieving accurate scene representation in complex scenarios while enhancing the detail-capturing capability.
The main contributions of this work are summarized as follows:
\begin{itemize}
	\item We propose an in-scattering model that integrates common light transports in real-world, and a new volume rendering method which is applicable to light scattering. 
	
	\item  Based on the proposed in-scattering model and volume rendering method, we design an enhanced deblur network, Is-NeRF, for complex real-world scenes.
	
	\item We develop a self-guided learning strategy of scattering directions and sampling intervals, achieving high-fidelity scene representation with finer geometric details.
	
	\item Extensive experiments on real-world datasets demonstrate Is-NeRF's effectiveness in handling motion-blurred images and complex light transport scenarios.
\end{itemize}

\section{Related Work}

\subsection{Neural Radiance Field}
Owing to its excellent 3D scene representation and novel view synthesis ability, NeRF has been applied to several areas, e.g. scene reconstruction \cite{li2021neural, sun2021neuralrecon,li2022neural}, 3D generative models \cite{niemeyer2021giraffe,schwarz2020graf}, relighting \cite{philip2021free,srinivasan2021nerv}, human-body reconstruction \cite{peng2021animatable,su2021nerf}, etc. However, there are few studies aiming at non-ideal conditions. Mip-NeRF \cite{barron2021mip} addresses the aliasing issue of ray sampling, and then Mip-NeRF360 \cite{barron2022mip} expand it to unbounded scenes. NeRF-W \cite{martin2021nerf} deals with appearance inconsistencies and transient objects in acquired images, HDR-NeRF \cite{huang2022hdr} simulates High Dynamic Range radiance to imitate the physical process of image acquisition. SeaThru-NeRF \cite{levy2023seathru} proposes a suitable architecture for scene learning in a scattering medium. 
On the other hand, many methods focus on lifting the rendering efficiency of NeRFs, which are mainly categorized into grid-based \cite{cao2023hexplane,chen2022tensorf,fridovich2023k,fridovich2022plenoxels} and hash-based \cite{muller2022instant,wang2020joint} methods. Despite these efforts, achieving real-time rendering of unbounded and complete scenes remains challenging. All these methods rely on accurate camera poses and high quality images. 

\subsection{Deblur Neural Radiance Fields}
In practice, training accurate neural radiance field from blurry images is more common. 
Deblur-NeRF \cite{ma2022deblur} explores a new area of training neural radiance fields from blurred images end-to-end by simulating blurring kernals to optimize potential sharp images during inference. Luthra et al. \cite{luthra2024deblur} propose to learn spatial varying blurring kernels. DP-NeRF \cite{lee2023dp} and Sharp-NeRF \cite{Lee_2024_WACV} enhance the deblurring quality by designing more efficient kernels. However, these methods prefer defocus blur. 

Several works have been proposed to deal with motion blur, gratifying results have been obtained for both image deblurring \cite{kupyn2019deblurgan,tao2018scale} and video deblurring \cite{su2017deep}. Haesol et al. \cite{park2017joint} jointly recover camera pose, dense depth map, and potentially sharp images from multi-view motion blurred images. 
MP-NeRF \cite{2024MP} and MBS-NeRF \cite{2025MBS} introduce depth and prior information. BAD-NeRF \cite{wang2023bad} improves the robustness in intense motion blurred images by explicitly modeling the physical causes of motion blur. Moreover, event-driven methods \cite{Qi_2023_ICCV, 10.1145/3664647.3680569, 2025Deblur} for event camera also gain success. ScatterNeRF \cite{ramazzina2023scatternerf} introduces attenuation coefficient to model light scattering in inclement weather conditions, and Aleth-NeRF \cite{cui2024alethnerf} integrates concealing field in volume rendering for low-light $\&$ over-exposure. NeRF-based methods employ traditional straight-line sampling, struggle to handle complex light propagation, leading to geometric ambiguities in real-world scenarios. New ideas are expected in this field.

\section{Modeling}
\subsection{In-scattering Lightpath Model}
In real-world scenarios with extensive light reflection, refraction, or transmission, the straight-line sampling strategy in standard volume rendering incorrectly interprets images formed by reflected or refracted rays as originating from objects positioned along straight paths, causing inconsistencies between reconstructed property and physical reality. This misguidance would be amplified by motion blurring, hindering convergence to accurate color and geometry.

\begin{figure}[!h]
	\centering
	\includegraphics[width=0.45\textwidth]{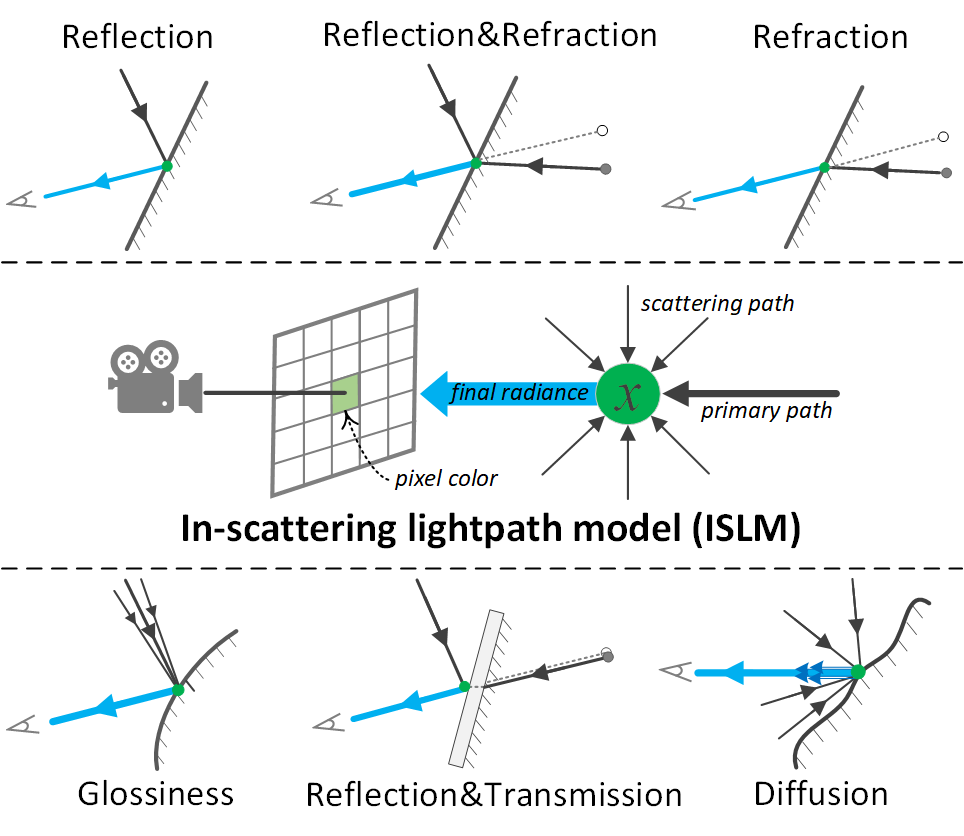}
	\caption{
		The in-scattering lightpath model (ISLM). It unifies six commonly seen phenomena.
	}
	\label{fig_islm}
\end{figure}

According to scattering theory, photons from any direction can scatter into the current lightpath and contribute to the final radiance, namely, in-scattering. By analyzing the daily scenarios, we summarize them into six ways of light transportation and unify them as in-scattering (Figure \ref{fig_islm} and Appendix A), where lights meet at a point on the medium or surface, with multiple scattered rays converging into the camera’s incident path to form a pixel. This In-scattering Lightpath Model (ISLM) decomposes the complex lightpath into two parts: primary lightpath and scattering lightpath. The sampling points along both parts can be learned and combined to render pixel color.

\begin{figure*}[]
	\centering
	\includegraphics[width=0.9\textwidth]{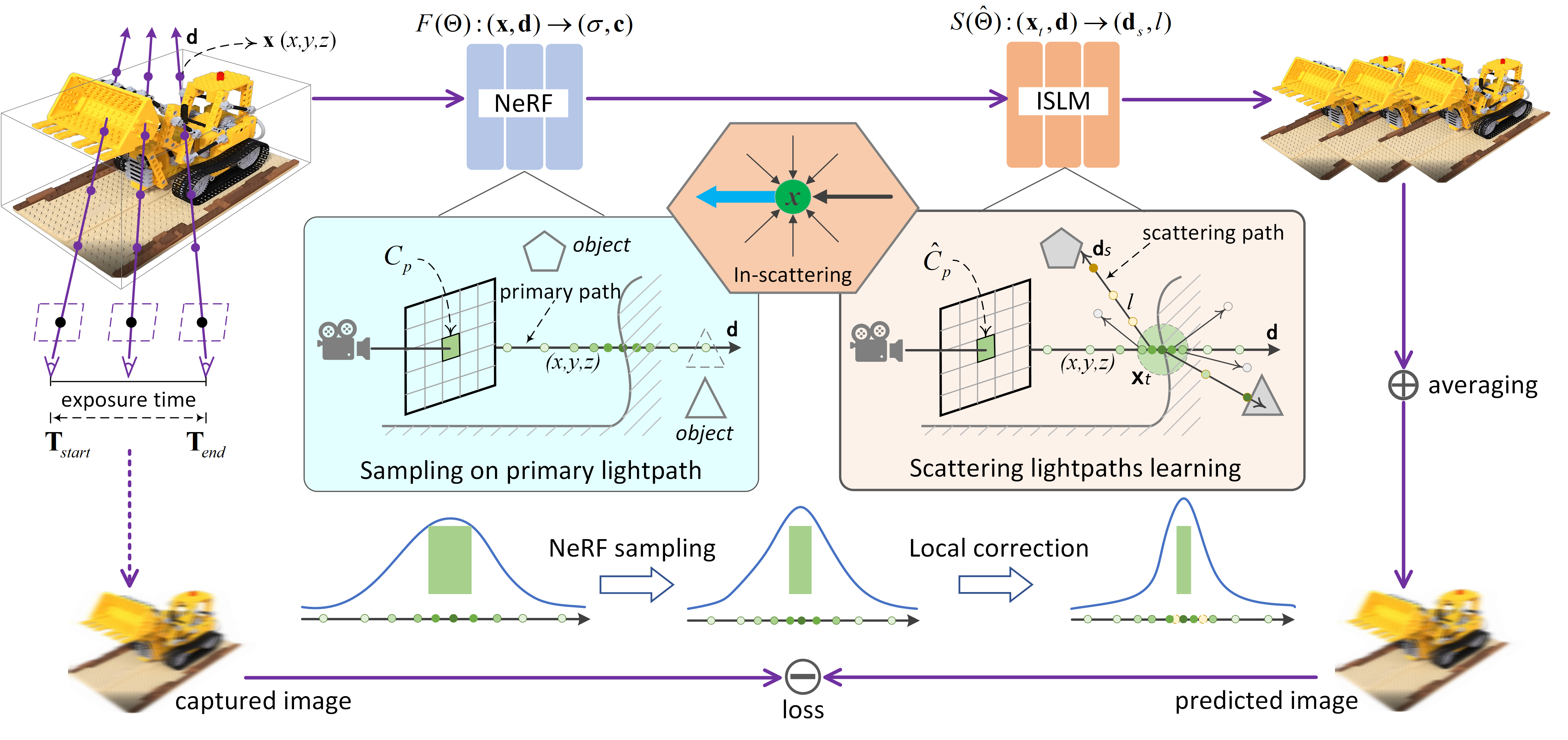}
	\caption{
		\textbf{The pipeline of Is-NeRF.} Given a set of motion blurred images, first we try to learn 3D representations along primary paths ($F(\Theta)$) and camera trajectories in exposure time. Then we propose an In-scattering Lightpath Model (ISLM) and design a scattering-aware volume rendering which are used to learn scattering representations ($S(\hat{\Theta})$ and to render virtual sharp images. During this process, the local density of sampling points is corrected to capture better details. The blurry image is synthesized by averaging those virtual images and used to construct photo-metric loss for network training.
	}
	\label{fig_network}
\end{figure*}

\subsection{Scattering-aware Volume Rendering}
With ISLM generates new sampling points in the scattering directions, the final color of a pixel is then jointly rendered by the volume densities and radiance contributions from all sampling points on both the primary and scattering lightpaths. The mathematical formulation of ISLM is defined as:
\begin{equation}
	S(\hat{\Theta}): (\mathbf{x}_t, \mathbf{d}) \rightarrow (\mathbf{d}_s, l)  
\end{equation}
This function maps the spatial coordinates $\mathbf{x}_t$ of a sampling point on the primary path $\mathbf{d}$ to its scattering path $\mathbf{d}_s$ and the sampling interval $l$. Unlike NeRF’s sampling strategy, ISLM directly controls the sampling interval (see Figure \ref{fig_network}) to regulate the length of scattering path. The coordinates of sampling points along scattering path are computed by Eq. (\ref{eq_samples}), where $\mathbf{x}_{tj}$ denotes the $j$-th sample on scattering path $\mathbf{d}_s$ originated from $\mathbf{x}_t, N_t$ is the number of sampling points.
\begin{equation}\label{eq_samples}
	\mathbf{x}_{tj} = \mathbf{x}_t + j \cdot l \cdot \mathbf{d}_s, \quad j \in \{1, 2, \dots, N_t\}  
\end{equation}

Then, the final pixel color $\hat{C}_p$ integrates contributions from both the primary and scattering paths. The equation of volume rendering is extended as follows:
\begin{equation}\label{eq_vr}
	\hat{C}_p = C_p + \sum_{k=1}^K \sum_{i=1}^{N_k} T_{ki} (1- \exp(-\sigma_{ki} l_{k} )) \mathbf{c}_{ki} 
\end{equation}
\begin{equation}
	T_{ki} = \exp\left(-\sum_{j=1}^{i-1} \sigma_{kj} \delta_{kj} \right) 
\end{equation}
Here, $C_p$ means the color contribution along primary path.
$K$ and $N_k$ are parameters controlling the number of scattering paths and sampling points per path, respectively.
$\sigma_{ki}$ and $\mathbf{c}_{ki}$ represent the volume density and color at the $i$-th sample on the $k$-th scattering path.
$T_{ki}$ acts as a path validity mask: for incorrectly predicted paths, $T_{ki}$ suppresses their contribution to the final color, ensuring ISLM converges to physically accurate lightpaths.

\section{Framework}

By integrating ISLM, we proposes a new framework, named Is-NeRF, to deal with blurred images. It jointly optimizes NeRF parameters, ISLM parameters and camera trajectories, progressively refining ideal straight-line sampling into physically accurate scattering lightpaths to adapt various real-world scenes. Figure \ref{fig_network} illustrate the overall structure.

\subsection{Network pipeline}
There are four main steps:
(1) Camera pose interpolation. Linearly interpolate between the initial pose $\mathbf{T}_{\text{start}}$ and the final pose $\mathbf{T}_{\text{end}}$ in $\textbf{SE}(3)$ space to generate sampling moments and corresponding virtual camera poses (see Appendix A).
(2) Primary path sampling. Follow NeRF's coarse-to-fine strategy to predict color and density ($F(\Theta)$) at spatial sampling points along primary lightpath.
(3) ISLM-guided scattering learning. Taking primary path and the chosen sampling point as input, ISLM then learns scattering directions $\mathbf{d}_s$ and sampling intervals $l$ grow from the primary point ($S(\hat{\Theta})$).
(4) Blur synthesis and optimization. Render virtual sharp images for each sampling moment by Eq. (\ref{eq_vr}), average them to predict motion-blurred images, and minimize the photometric error between predicted and input blurred images. The NeRF parameters $\Theta$, ISLM parameters $\hat{\Theta}$ , as well as $\mathbf{T}_{\text{start}}$ and $\mathbf{T}_{\text{end}}$ are jointly optimized.

\subsection{Adaptive Sampling of ISLM}
Determining scattering lightpaths and their sampling points is critical to ISLM and the network. Theoretically, multiple scattering paths should originate from the intersection of the primary lightpath and object surface, with sampling points generated according to NeRF's sampling principles. By parameter adjustments, the network would gradually converge to accurate paths. However, selecting sole intersection point in practice is overly idealized (sampling points on primary path themselves are statistically estimated) and contradicts the principles of volume rendering, thereby can not guarantee robustness. Inspired by ray-tracing optimization, this work constructs scattering paths from adjacent sampling points along the primary path and assigns each point one scattering path to collectively simulate light scattering. There are three key things:
(1) The quantity of scattering paths. Five primary points are selected to grow scattering paths. which jointly simulate the scattering locations and global effects (see Figure \ref{fig_network}).
(2) Scattering direction. By fixing the count of primary points and scattering paths, the network adaptively learns the direction $\textbf{d}_s$ of each path, ensuring convergence of all paths.
(3) Sampling along scattering path. To balance training efficiency and precision, we employ fixed-number equidistant sampling on scattering paths, where interval $l$ is another learnable parameter.

\subsection{Local Density Correction}
NeRF's hierarchical sampling strategy relies heavily on the initial sampling of coarse network. When small objects occupy the space, most samples may fall off the object and cause the bandwidth of peak of probability density to expand. This forces the network to adjust parameters of adjacent points to approximate the true geometry. The intended probability density peak is averaged across neighboring samples, leading to inaccurate geometric representations.
Is-NeRF addresses this issue by generating network-controlled scattering lightpaths through ISLM. These scattering paths increase the probability of sampling points hitting small objects. Moreover, the intervals along scattering paths are learned by the network and jointly optimized with other parameters, enabling adaptive adjustments. This strategy enhances Is-NeRF’s perception of fine structures compared to prior methods, as validated by experimental results. 

\subsection{Loss Function}
Given  $M$  motion-blurred images, our method can jointly optimize the parameters of ISLM ($\hat{\Theta}$), NeRF ($\Theta$), and the camera motion trajectories (i.e., $\mathbf{T}_{\text{start}}$ and $\mathbf{T}_{\text{end}}$) by minimizing the photometric loss which is formulated as:
\begin{equation}
	\mathcal{L} = \sum_{m=1}^{M} \sum_{\mathbf{x}} \left\| \hat{B}_m(\mathbf{x}) - B^{\text{gt}}_m(\mathbf{x}) \right\|_2^2  
\end{equation}
where $\hat{B}_m(\mathbf{x})$ is the motion-blurred image predicted by Is-NeRF, and $B^{\text{gt}}_m(\mathbf{x})$ denotes the captured blurring image. The following Jacobian matrices are required to update parameters via stochastic gradient descent. 

\begin{equation}
	\frac{\partial \mathcal{L}}{\partial \hat{\Theta}}=\sum_{m=1}^{M}\frac{\partial \mathcal{L}}{\partial B_m(\mathbf{x})} \cdot \frac{1}{n} \sum_{i=1}^{n} \frac{\partial B_m(\mathbf{x})}{\partial \hat{C}_i} \cdot \frac{\partial \hat{C}_i}{\partial \hat{\Theta}}
\end{equation}
\begin{equation}
	\frac{\partial \mathcal{L}}{\partial \Theta}=\sum_{m=1}^{M}\frac{\partial \mathcal{L}}{\partial B_m(\mathbf{x})} \cdot \frac{1}{n} \sum_{i=1}^{n} \frac{\partial B_m(\mathbf{x})}{\partial \hat{C}_i} \cdot \frac{\partial \hat{C}_i}{\partial \Theta}
\end{equation}
\begin{equation}
	\frac{\partial \mathcal{L}}{\partial \mathbf{T}_{\text{start}}}=\sum_{m=1}^{M}\frac{\partial \mathcal{L}}{\partial B_m(\mathbf{x})} \cdot \frac{1}{n} \sum_{i=1}^{n} \frac{\partial B_m(\mathbf{x})}{\partial \hat{C}_i} \cdot \frac{\partial \hat{C}_i}{\partial \mathbf{T}_{\text{start}}}
\end{equation}
\begin{equation}
	\frac{\partial \mathcal{L}}{\partial \mathbf{T}_{\text{end}}}=\sum_{m=1}^{M}\frac{\partial \mathcal{L}}{\partial B_m(\mathbf{x})} \cdot \frac{1}{n} \sum_{i=1}^{n} \frac{\partial B_m(\mathbf{x})}{\partial \hat{C}_i} \cdot \frac{\partial \hat{C}_i}{\partial \mathbf{T}_{\text{end}}}
\end{equation}

\section{Experimental Evaluation}
We conduct parameter evaluations, comparative evaluations, and ablation studies on two datasets: a public motion-blurred dataset provided by Deblur-NeRF \cite{ma2022deblur} and a self-captured dataset. All experiments are conducted on a NVIDIA RTX3090 GPU (24GB). The rendered images are evaluated in three metrics: PSNR, SSIM, and LPIPS.

\subsection{Parameter Evaluations}
\label{exp_1}
\begin{figure}[!h]
	\centering
	\includegraphics[width=0.48\textwidth]{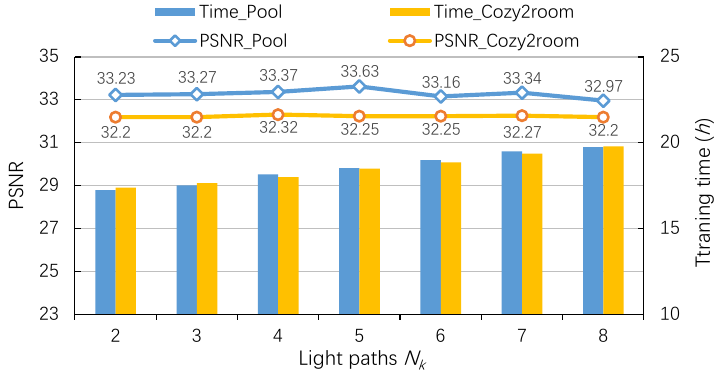}
	\caption{
		The PSNR and training time under varying number of scattering lightpaths $N_k$.
	}
	\label{fig:fig_3}
\end{figure}

The number of primary points initiating scattering $N_k$ critically influence Is-NeRF’s performance.
Theoretically, larger $N_k$ corresponds to finer detail modeling and better handling of light scattering. However, excessive $N_k$ hinders convergence of model and reduces efficiency. Conversely, insufficient $N_k$ fails to adequately model light scattering, degrading performance.
Figure \ref{fig:fig_3} shows the PSNR and training time for scene ‘Pool’ and ‘Cozy2Room’ across varying $N_k$. Results reveal that $N_k = 5$, originating scattering paths from five adjacent primary points, balances accuracy and efficiency, while excessive $N_k$ only inflates training time without improving image quality.

\begin{table}[!h]
	\centering
	\small
	\resizebox{0.48\textwidth}{!}{
	\begin{tabular}{l|ccc|ccc}
		\toprule
		\multirow{2}{*}{\textbf{Scene}}& 
		\multicolumn{3}{c}{\textbf{5 rays at 1 point}} & 
		\multicolumn{3}{c}{\textbf{1 ray at each of 5 points}} \\
		\cmidrule(lr){2-4} \cmidrule(l){5-7}
		& PSNR$\uparrow$ & SSIM$\uparrow$ & LPIPS$\downarrow$ & PSNR$\uparrow$ & SSIM$\uparrow$ & LPIPS$\downarrow$ \\
		\midrule
		Cozy2room  & 28.77 & 0.8598 & 0.1443 & \textbf{32.25} & \textbf{0.9162} & \textbf{0.0515} \\
		Pool       & 27.83 & 0.7128 & 0.3539 & \textbf{33.53} & \textbf{0.8962} & \textbf{0.0749} \\
		Bottle     & 23.19 & 0.8240 & 0.3293 & \textbf{23.89} & \textbf{0.9403} & \textbf{0.0836} \\
		LakeCorner & 25.54 & 0.6926 & 0.4091 & \textbf{27.14} & \textbf{0.7780} & \textbf{0.3013} \\
		\bottomrule
	\end{tabular}
	}
	\caption{Quantitative comparison of two scattering designs. It is evident that the second design prevails in all metrics.}
	\label{tab:scattering}
\end{table}

Additionally, we compare two scattering designs: 1) emitting five rays from a single primary sampling point, versus 2) emitting one scattering ray from each of five adjacent primary points. The results in Table \ref{tab:scattering} show that emitting rays from five consecutive primary points outperforms the single-point multi-ray style (Intuitive results given in Appendix B also support this conclusion). This aligns with the principle of volume rendering that more scattering origins increase the likelihood of capturing true scattering locations. Based on these findings, we adopt the second design, prioritizing scattering path diversity over single-point multiplicity.

\begin{table*}[]
	\centering
	\definecolor{first}{RGB}{255,140,140}
	\definecolor{second}{RGB}{255,180,180}
	\definecolor{third}{RGB}{255,220,220}
	\newcommand{\ca}{\cellcolor{first}}
	\newcommand{\cb}{\cellcolor{second}}
	\newcommand{\cc}{\cellcolor{third}}
	\resizebox{\textwidth}{!}{
		\begin{tabular}{l|ccc|ccc|ccc|ccc|ccc}
			\toprule
			\multirow{2}{*}{\textbf{Method}} & \multicolumn{3}{c|}{\textbf{Cozy2room}}& \multicolumn{3}{c|}{\textbf{Factory}} & \multicolumn{3}{c|}{\textbf{Pool}} &\multicolumn{3}{c|}{\textbf{Tanabata}} &\multicolumn{3}{c}{\textbf{Trolley}}  \\
			& PSNR$\uparrow$ & SSIM$\uparrow$ & LPIPS$\downarrow$ & PSNR$\uparrow$ & SSIM$\uparrow$ & LPIPS$\downarrow$ & PSNR$\uparrow$ & SSIM$\uparrow$ & LPIPS$\downarrow$ & PSNR$\uparrow$ & SSIM$\uparrow$ & LPIPS$\downarrow$ & PSNR$\uparrow$ & SSIM$\uparrow$ & LPIPS$\downarrow$ \\
			\midrule
			NeRF (2021) & 25.66 & 0.7941 & 0.2288 & 19.32 &0.4563 &0.5304 & 30.45 & 0.8354 & 0.1932 & 22.22 & 0.6807 & 0.3653 & 21.25 & 0.6370 & 0.3633  \\
			MPR (2021)& 29.90 & 0.8862 & 0.0915 & 25.07 & 0.6940 & 0.2409 & \cb 33.28 & \cb 0.8938 & 0.1290 & 22.60 & 0.7203 & 0.2507 & 26.24 & 0.8356 & 0.1762 \\
			PVD (2021) & 28.06 & 0.8443 & 0.1515 & 24.57 & 0.6877 & 0.3150 & 30.38 & 0.8393 & 0.1977 & 22.54 & 0.6872 & 0.3351 & 24.44 & 0.7746 & 0.2600 \\
			SRN-Deblur (2018) & 29.47 & 0.8759 & 0.0950 & 26.54 & 0.7604 & 0.2404 & 32.94 & 0.8847 & 0.1045 & 23.20 & 0.7274 & 0.2438 & 25.36 & 0.8119 & 0.1618 \\
			Deblur-NeRF (2022) & 25.96 & 0.7979 & 0.1024 & 23.21 & 0.6487 & 0.2618 & 31.21 & 0.8518 & 0.1382 & 22.46 & 0.6946 & 0.2455 & 24.94 & 0.7923 & 0.1766 \\
			DP-NeRF (2023) & 30.77 & 0.9020 & 0.0584 & 27.69 & 0.8328 &  0.1847 & \cc 33.22 & \cc 0.8922 & 0.0954 &  25.27 & 0.7973 & 0.1779 & 26.99 & 0.8413 &  0.1312 \\
			Sharp-NeRF (2024) &25.65 &0.7994 &0.2322 &19.78 &0.4498 &0.4596 &30.53 &0.8520 &0.1845 &18.12 &0.4609 &0.4633 &21.67 &0.6992 &0.3073  \\
			MP-NeRF (2024) & \ca 32.49 & \ca 0.9296 & \cb 0.0368 & \cc28.45 & \cc 0.8497 & \cc 0.1629 &  31.30 & 0.8641 & 0.1029 & \cb 27.82 &  \ca 0.8821 & \cb 0.0872 & \cc 28.67 &\cc 0.8832 & \cc 0.0854 \\
			BAD-NeRF (2023) & \cc 32.15 &  0.9139 & 0.0527 & \cb 32.13 & \cb 0.9106 & \cb 0.1219 & 32.22 & 0.8600 &\cc 0.0908 & \cc 27.76 &  0.8600 & 0.1195 & \cb 29.25 & \cb 0.8892 & \cb 0.0833 \\
			\hline
			3DGS (2023) &  25.59 &  0.8076 &  0.1645 & 18.11 & 0.4179 & 0.4958 & 25.63 & 0.6326 &0.2632 & 21.35 & 0.6686 & 0.3235 & 20.56 & 0.6257 & - \\
			Deblurring-3DGS (2024) &  31.45 & \cb 0.9222 &\ca  0.0367 & 24.01 & 0.7333 & 0.2326 & 31.87 & 0.8829 &\cb 0.0751 & 27.01 &\cb 0.8807 &\ca 0.0785 & 26.88 & 0.8710 & - \\
			\hline
			\textbf{Is-NeRF (ours)}& \cb 32.25 & \cc 0.9162 & \cc 0.0515 & \ca 32.31 & \ca 0.9154 & \ca 0.1181 & \ca 33.53 & \ca 0.8962 & \ca 0.0749 & \ca 27.87 & \cc 0.8614 & \cc 0.1191 & \ca 29.27 & \ca 0.8894 & \ca 0.0787 \\
			\bottomrule
		\end{tabular}
	}
	\caption{Quantitative results on Deblur-NeRF dataset. The data reveal that our Is-NeRF achieves the best performance, especially on scene 'Pool' with extensive scatterings. MP-NeRF excels on 'Cozy2room', and BAD-NeRF show its deblur ability.}
	\label{tab:experiment1}
\end{table*}

\begin{figure*}[!h]
	\centering
	\includegraphics[width=0.95\textwidth]{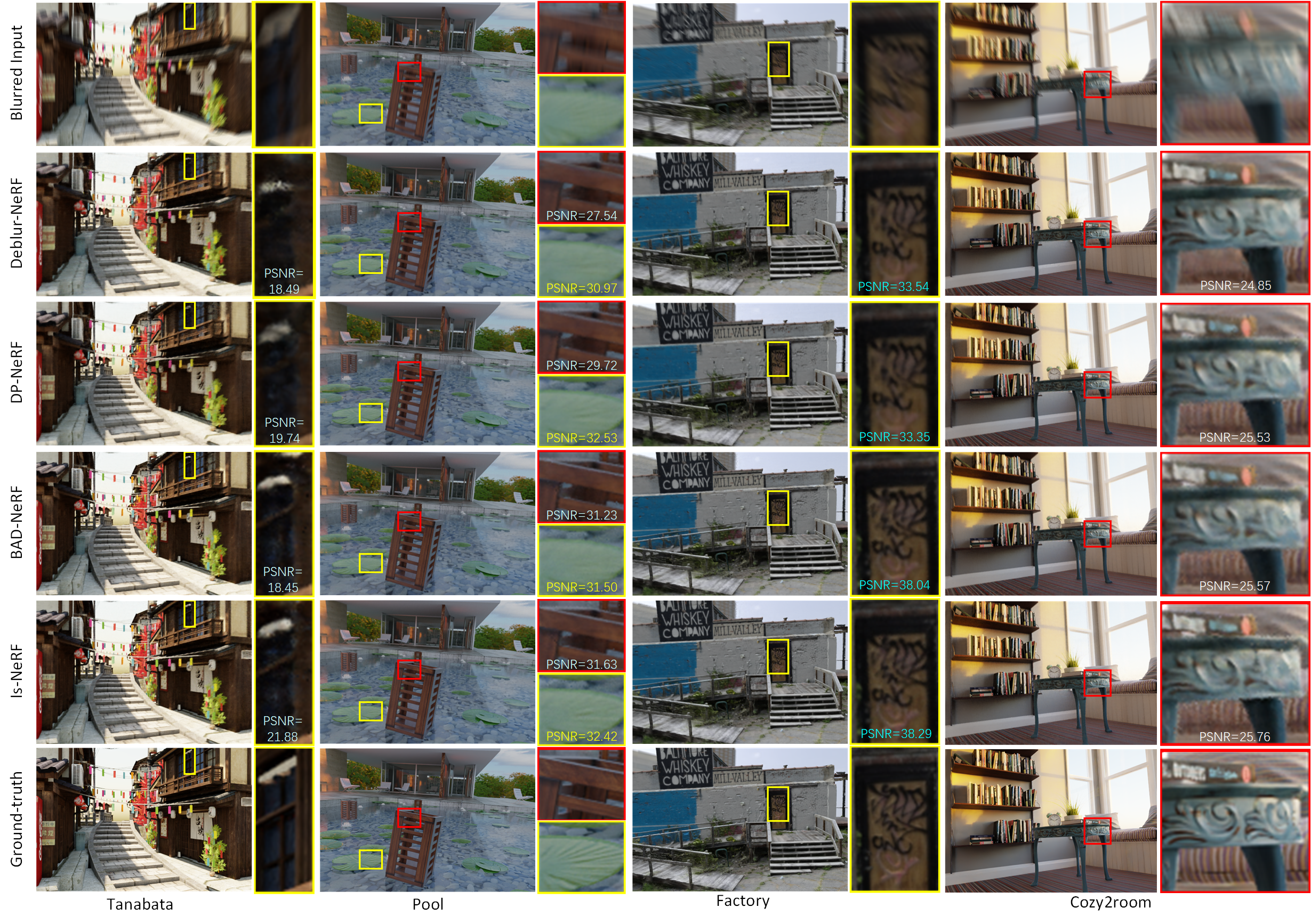}
	\caption{
		Intuitive results on Deblur-NeRF dataset, with \textbf{local regions enlarged and PSNR given.} The qualitative results show the advantage of our method in detail capturing.
	}
	\label{fig:visual1}
\end{figure*}

\begin{table*}[!h]
	\centering
	\small
	\definecolor{first}{RGB}{255,140,140}
	\definecolor{second}{RGB}{255,180,180}
	\definecolor{third}{RGB}{255,220,220}
	\newcommand{\ca}{\cellcolor{first}}
	\newcommand{\cb}{\cellcolor{second}}
	\newcommand{\cc}{\cellcolor{third}}
	\resizebox{\textwidth}{!}{
		\begin{tabular}{l|ccc|ccc|ccc|ccc} 
			\toprule
			\multirow{2}{*}{\textbf{Method}}& 
			\multicolumn{3}{c|}{\textbf{Bottle}} & \multicolumn{3}{c|}{\textbf{LakeCorner}} & 
			\multicolumn{3}{c|}{\textbf{Car}} & \multicolumn{3}{c}{\textbf{Table}} \\
			& {PSNR$\uparrow$} & {SSIM$\uparrow$} & {LPIPS$\downarrow$} & {PSNR$\uparrow$} & {SSIM$\uparrow$} & {LPIPS$\downarrow$} & {PSNR$\uparrow$} & {SSIM$\uparrow$} & {LPIPS$\downarrow$} & {PSNR$\uparrow$} & {SSIM$\uparrow$} & {LPIPS$\downarrow$} \\
			\midrule
			NeRF (2021)   & 20.72 & 0.7582 & 0.4113 & 11.36 & 0.4067 & 0.7899  & 28.36 & 0.8489 & 0.2403 & 25.39 & 0.7887 & 0.2600 \\
			SRN-Deblur (2018)  & \ca 29.86 & \cb 0.9288 & \cc 0.1107 & \cc 25.55 & \cb 0.7150 & 0.3855 & \cc 30.20 & 0.8790 & 0.1743 & 26.10 & 0.8103 & 0.2242 \\
			Deblur-NeRF (2022)  & 18.72 & 0.8552 & 0.1788 & 19.88 & 0.5833 & \cc 0.3295 & 29.50 & 0.8806 & 0.1203 & 28.63 & 0.8871 & 0.1078 \\
			DP-NeRF  (2023)   & 22.67 & \cc 0.9038 & \cb 0.1047 & 18.55 & 0.5390 & 0.3456 & 29.99 & \cc 0.8974 & \ca 0.0853 & \cc 29.02 & \cc 0.8949 & 0.1052  \\
			MP-NeRF (2024) & 21.93 &0.8802 &0.1211 &18.56 &0.5635 &0.3574 &29.65 &0.8830 &0.1012 &29.00 &0.8847 &\cc 0.0921  \\
			BAD-NeRF  (2023)   & \cc 23.32 & 0.7857 & 0.2203 & \cb 26.56 & \cc 0.6892 & \cb 0.3048 & \cb 31.86 & \cb 0.9041 & \cc 0.0912 & \cb 30.95 & \cb 0.9093 & \cb 0.0710\\
			\hline
			\textbf{Is-NeRF (ours)} & \cb 23.89 & \ca 0.9403 & \ca 0.0836 & \ca 27.14 & \ca 0.7780 & \ca 0.3013 & \ca 31.94 & \ca 0.9048 & \cb 0.0900 & \ca 31.20 & \ca 0.9142 & \ca 0.0652 \\
			\bottomrule
		\end{tabular}
	}
	\caption{Quantitative results on self-captured dataset with complex light propagation. The results reveal that our method achieves the best deblur ability in real-world scenarios.}
	\label{tab:exp2}
\end{table*}

\begin{figure*}[!h]
	\centering
	\includegraphics[width=0.95\textwidth]{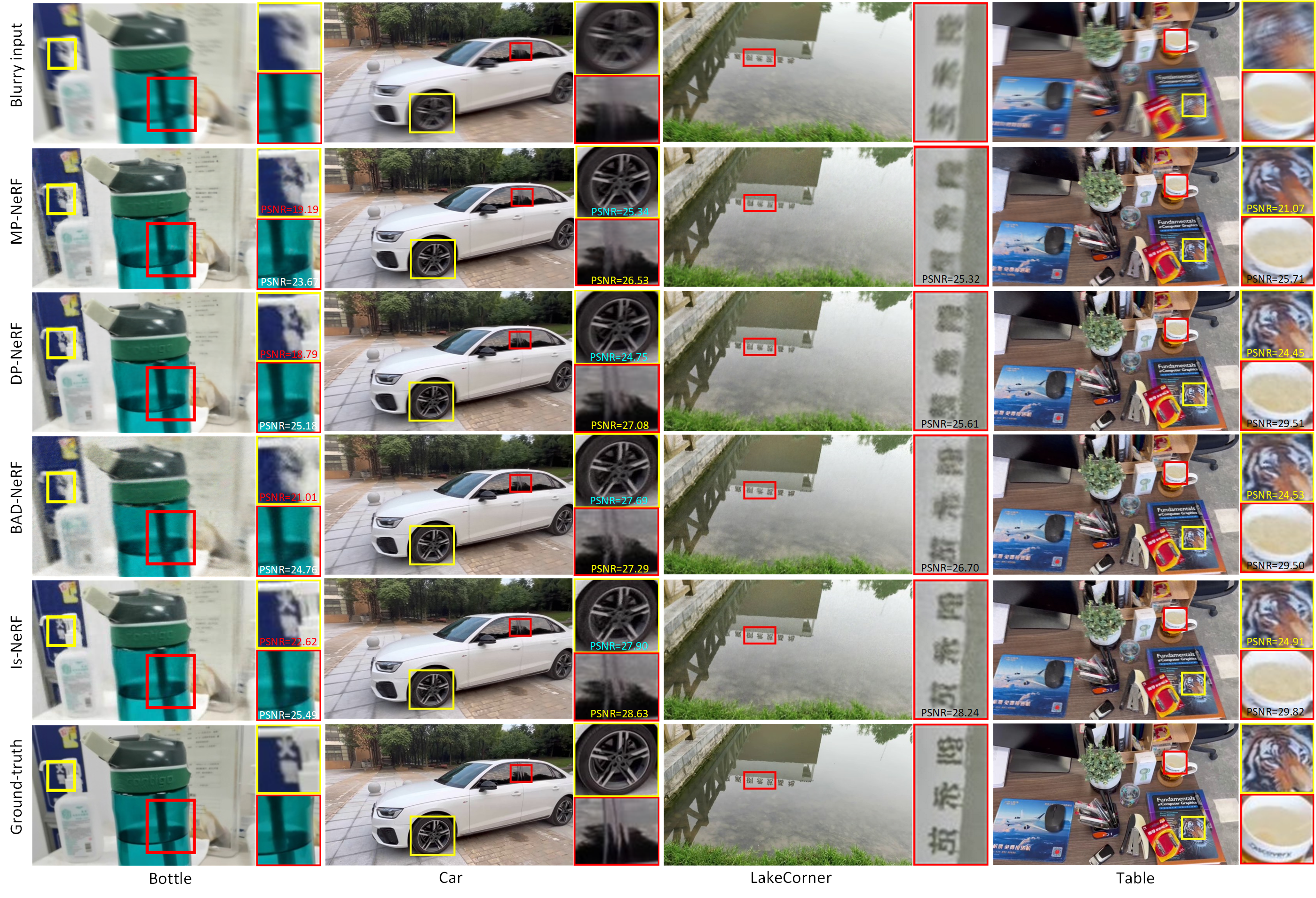}
	\caption{
		Intuitive results on self-captured dataset, with \textbf{local regions enlarged and PSNR given}. The results show that our method outperforms other method on real-world complex scenarios with extensive scatterings.
	}
	\label{fig:exp2}
\end{figure*}

\subsection{Comparison with SOTA methods}
We compare the proposed method with SOTA deblurring NeRF approaches (SRN-Deblur \cite{tao2018scale}, PVD \cite{son2021recurrent}, MPR \cite{zamir2021multi}, Deblur-NeRF \cite{ma2022deblur}, DP-NeRF \cite{lee2023dp}, Sharp-NeRF \cite{Lee_2024_WACV}, MP-NeRF \cite{2024MP}, baseline BAD-NeRF \cite{wang2023bad}), as well as 3DGS \cite{kerbl20233d} and Deblurring-3DGS \cite{lee2024deblurring3dgs}, emphasizing its ability to learn local details and scattering lightpaths.

\begin{table*}[!h]
	\centering
	\small
	\sisetup{table-format=2.4, table-number-alignment=center}
	\definecolor{first}{RGB}{255,140,140}
	\definecolor{second}{RGB}{255,180,180}
	\definecolor{third}{RGB}{255,220,220}
	\newcommand{\ca}{\cellcolor{first}}
	\newcommand{\cb}{\cellcolor{second}}
	\newcommand{\cc}{\cellcolor{third}}
	\resizebox{\textwidth}{!}{
		\begin{tabular}{@{} l cc *{4}{S[table-format=2.2] S[table-format=1.4] S[table-format=1.4]} @{}}
			\toprule
			\multirow{2}{*}{\textbf{Method}} & 
			\multicolumn{2}{c}{I\textbf{SLM}} &
			\multicolumn{3}{c}{Cozy2room} & \multicolumn{3}{c}{Factory} & \multicolumn{3}{c}{Pool} &\multicolumn{3}{c}{Tanabata} \\
			\cmidrule(lr){2-3} \cmidrule(lr){4-6} \cmidrule(lr){7-9} \cmidrule(lr){10-12} \cmidrule(l){13-15}
			& {Training} & {Rendering}
			& {PSNR$\uparrow$} & {SSIM$\uparrow$} & {LPIPS$\downarrow$} 
			& {PSNR$\uparrow$} & {SSIM$\uparrow$} & {LPIPS$\downarrow$}
			& {PSNR$\uparrow$} & {SSIM$\uparrow$} & {LPIPS$\downarrow$} 
			& {PSNR$\uparrow$} & {SSIM$\uparrow$} & {LPIPS$\downarrow$}\\
			\midrule
			Baseline & $\times$ & $\times$ & 32.15 &  0.9139 &  0.0527 &  32.13 &  0.9106 & 0.1219 & 33.10 & 0.8864 & 0.0802 & 27.76 & 0.8600 & 0.1195 \\
			Is-NeRF &$\times$ & $\checkmark$ & 13.24 & 0.4270 & 0.6393 & 14.73 & 0.6393 & 0.4914 & 32.97 & 0.8950 & 0.0781 & 18.62 & 0.7688 & 0.1896 \\
			Is-NeRF & $\checkmark$ & $\times$ & 29.31 & 0.8958 & 0.0873 & 29.96 & 0.8996 & 0.1288 & 33.51 & \textbf{0.8962} & \textbf{0.0749} & 27.86 & \textbf{0.8614} & \textbf{0.1192} \\
			\textbf{Is-NeRF} & $\checkmark$  & $\checkmark$  &\textbf{32.25} & \textbf{0.9162} & \textbf{0.0515} & \textbf{32.31} & \textbf{0.9154} & \textbf{0.1181} & \textbf{33.53} & \textbf{ 0.8962} & \textbf{0.0749} & \textbf{27.87} & \textbf{0.8614} & 0.1191\\
			\hline
		\end{tabular}
	}
	\resizebox{\textwidth}{!}{
		\begin{tabular}{@{} 
				l 
				cc 
				*{4}{S[table-format=2.2] S[table-format=1.4] S[table-format=1.4]} 
				@{}
			}
			\multirow{2}{*}{\makecell[l]{\textbf{Method}}} & 
			\multicolumn{2}{c}{\textbf{ISLM}} &
			\multicolumn{3}{c}{Bottle} & \multicolumn{3}{c}{LakeCorner} & \multicolumn{3}{c}{Car} & \multicolumn{3}{c}{Table} \\
			\cmidrule(lr){2-3} \cmidrule(lr){4-6} \cmidrule(lr){7-9} \cmidrule(lr){10-12} \cmidrule(l){13-15} 
			& {Training} & {Rendering}
			& {PSNR$\uparrow$} & {SSIM$\uparrow$} & {LPIPS$\downarrow$}
			& {PSNR$\uparrow$} & {SSIM$\uparrow$} & {LPIPS$\downarrow$} 
			& {PSNR$\uparrow$} & {SSIM$\uparrow$} & {LPIPS$\downarrow$}
			& {PSNR$\uparrow$} & {SSIM$\uparrow$} & {LPIPS$\downarrow$} \\
			\midrule
			Baseline & $\times$ & $\times$  & 23.32 & 0.7857 & 0.2203 & 26.56 & 0.6892 & 0.3048 & 31.86 & 0.9041 & 0.0912 & 30.95 & 0.9093 & 0.0710 \\
			Is-NeRF & $\times$ & $\checkmark$  & 22.64 & 0.8603 & 0.2515 & 26.61 & 0.7062 & 0.3053 & 31.90 & 0.8979 & 0.0912 & 31.08 & 0.8961 & 0.0673 \\
			Is-NeRF & $\checkmark$ & $\times$ & 23.67 & 0.8313 & 0.0924 & 27.02 & 0.7075 &\textbf{0.3005} & 31.36 & 0.8936 & 0.0977 & \textbf{31.20} & 0.8964 & 0.0655 \\
			\textbf{Is-NeRF} & $\checkmark$  & $\checkmark$  & \textbf{23.89} & \textbf{0.9403} & \textbf{0.0836} & \textbf{27.14} & \textbf{0.7780} & 0.3013 & \textbf{31.94} & \textbf{0.9048} & \textbf{0.0900} & \textbf{31.20} & \textbf{0.9142} & \textbf{0.0652} \\
			\bottomrule
		\end{tabular}
	}
	\caption{The quantitative results of ablation studies of ISLM module to the proposed framework.}
	\label{tab:performance}
\end{table*}

\subsubsection{Results on Deblur-NeRF Dataset}
The public Deblur-NeRF dataset contains five scenes: four (Cozy2Room, Factory, Tanabata, and Trolley) with typical indoor/outdoor diffuse reflections, and one (Pool) with light reflection and refraction. All methods take blurred images as input. Quantitative results are summarized in Table \ref{tab:experiment1}.
Results reveal that Is-NeRF achieves significant improvements over single-view deblurring approaches (SRNDeblur, PVD, MPR) which rely on limited information from single view. Compared with Deblur-NeRF \cite{ma2022deblur}, our method outperforms owing to its physics-aware motion blur synthesis. Deblur-NeRF neglects the critical role of camera trajectories in motion blur formation. MP-NeRF \cite{2024MP} shows great potential but lacks of adaptability. By introducing scattering paths and local density correction under the scattering-aware volume rendering, Is-NeRF enhances learning ability of subtle geometric details comparing to baseline BAD-NeRF \cite{wang2023bad}. This leads to consistent performance gains in `Cozy2Room', `Factory', `Tanabata', and `Trolley'. For scene 'Pool' (rich in reflection and refraction), the improvement is particularly pronounced, validating the ISLM’s ability to model scattering light transport and correct errors from straight-path sampling. 

We also intuitively show the results in Figure \ref{fig:visual1}. Take ‘Pool’ as example, red box highlights large objects without light scattering, where Is-NeRF performs similarly to BAD-NeRF. Yellow box marks water area with more light reflection and refraction, where Is-NeRF outperforms BAD-NeRF, producing sharper and more detailed images. For other scenes with fine textures and geometric structures, Is-NeRF also achieves superior results through local volume density correction. These visual comparisons validate the effectiveness of our scattering-aware rendering.

\subsubsection{Results on self-captured dataset}
To further validate the advantages of the proposed Is-NeRF, we constructs a dataset featuring complex light transport by capturing real-life scenes. The comparative results are recorded in Table \ref{tab:exp2}. 
The results show that our Is-NeRF outperforms comparison methods across all three metrics, except that DP-NeRF \cite{lee2023dp} attains the best LPIPS metric for the scene ‘Car’. Deblur-NeRF \cite{ma2022deblur} and DP-NeRF exhibit significantly weaker performance in scenes with substantial refraction and reflection (Bottle, LakeCorner, Car) compared to other methods. For the indoor scene ‘Table’, SRN-Deblur and Deblur-NeRF produce inferior results. BAD-NeRF, which directly adopts NeRF's renderer and straight-line sampling, fails to adapt realistic light propagation. Consequently, it struggles to learn accurate geometric information in scenes with extensive light scattering, also validated by the  qualitative results.

To intuitively express Is-NeRF’s enhancements in detail learning and scattering lightpath representing, we present image synthesis results in Figure \ref{fig:exp2}, highlighting local regions and the PSNR values. Scene ‘Car’ contains reflections from glass surfaces and car paint, while ‘LakeCorner’ features large-scale regions with fraction and reflection. Our method demonstrate superior representation of such scenarios, recovering finer local details and achieving higher local PSNR comparing to other methods. For ‘Table’ scene, which includes glossy highlights (e.g., book cover) and refraction, our method also delivers the best visual representation. Scene ‘Bottle’ contains image textures and transmission effects. It is evident that Is-NeRF captures more details compared to BAD-NeRF. Additionally, Is-NeRF excels in modeling light transmission: the straw inside, objects behind the bottle, and shadow contours on the table are rendered with significantly clearer definition. 

The above results demonstrate that the proposed in-scattering lightpath model and the extended volume rendering handles sophisticated light transport effectively and can recover sharper scene details. This approach robustly adapts to diverse real-world scenarios, showcasing superior generalizability and applicability.

\subsection{Ablation Studies}
To investigate ISLM’s contribution to Is-NeRF, we conduct ablation studies. Removing ISLM from training implies disabling scattering path learning and local density correction, while removing ISLM from rendering retains scattering paths learned during training but performs standard volume rendering. If learning scattering lightpaths is unnecessary, removing ISLM module will not make noticeable difference.

As the data in Table \ref{tab:performance} reveals, removing the ISLM module significantly degrades synthesis quality on Deblur-NeRF dataset. For`Cozy2Room', `Factory', and `Tanabata', removing ISLM during training causes substantial performance degradation. For 'Pool' and self-captured scenes, the performance also declines moderately, though the model's overall deblurring capability remains superior to the baseline BAD-NeRF. Removing ISLM during rendering has a slightly smaller impact, with similar trends observed across both datasets. Intuitive comparisons are given in Appendix C.
These results reveal that ISLM plays a decisive role during the training: it corrects sampling rays to align with the true lightpaths and enables Is-NeRF to learn sufficient geometric information via local density correction. The promotion benefits from ISLM in training outweigh the loss caused by disabling it during rendering. This is critical for scenarios with rich reflection or refraction, where ISLM’s ability to gain physically consistent sampling is indispensable.


\section{Conclusion}

This work presents a new neural radiance field framework Is-NeRF to learn finer 3D representations of real-world scenes with sophisticated lightpaths from motion blurred images. We unify six daily seen light transport phenomena with an in-scattering model and propose an extended scattering-aware volume rendering method. Based on the theoretical modeling, a network that can adaptively learn scattering paths to optimize scene representations is designed, with local density correction ability to obtain accurate geometric details. Extensive experiments on synthetic and self-captured datasets then verify its superiority and applicability in complex real-life scenarios. 

This work brings new thoughts to neural radiance field and volume rendering. In the future, we will try to strike a balance between accuracy and speed by utilizing more efficient architectures (such as Zip-NeRF, Instant-NGP) as backbone. Another promising direction is to extend the in-scattering model to 3D Gaussian Splatting related fields.

\bibliography{myref}

\section{Appendix A}
\subsection{Neural Radiance field}
As an implicit 3D scene representation method, NeRF centers on using a deep neural network to model the radiance intensity and density at each point in a scene, synthesizing images through volume rendering. Training NeRF typically requires multiple high-quality 2D images and their corresponding camera parameters. It maps the 3D coordinates of a point $\mathbf{x} = (x, y, z)$ and the viewing direction $\mathbf{d} = (\theta, \phi)$ to the view-invariant volume density $\sigma$ and the view-dependent color $\mathbf{c} = (r, g, b)$:
\begin{equation}
	F(\Theta): (\mathbf{x}, \mathbf{d}) \rightarrow (\sigma, \mathbf{c})  
\end{equation}

Subsequently, an approximate volume rendering method predicts the color $C_p$ of each pixel using the predicted color $\mathbf{c}_{i}$ and density $\sigma_{i}$ by the NeRF backbone network at each sampled point $\mathbf{r}_{i}$ along the ray. The coloring equation is expressed as Eq. \ref{eq_2}, where $T_i$ denotes the accumulated transmittance and $\delta_i$ is the sampling interval.
\begin{equation}
	\label{eq_2}
	C_p = \sum_{i=1}^N T_i (1 - \exp(-\sigma_i \delta_i) ) \mathbf{c}_i, 
\end{equation} 
\begin{equation}
	T_i = \exp\left(-\sum_{j=1}^{i-1} \sigma_j \delta_j\right) 
\end{equation} 

\subsection{BAD-NeRF}
Unlike existing works (e.g., Deblur NeRF ) that treat blurred images as the convolution of sharp images with blur kernels, BAD-NeRF analyzes the physical formation process of motion blur. It posits that relative motion between objects and the camera during exposure time causes variations in captured radiance distributions across moments, ultimately manifesting as motion blur. The mathematical model can be approximated as:
\begin{equation}
	B(\mathbf{x}) \approx \frac{1}{n} \sum_{t=1}^n I_t(\mathbf{x})
\end{equation}
where  $B(\mathbf{x})$  represents the motion-blurred image at pixel coordinate  $\mathbf{x} \in \mathbb{R}^2 $, and  $I_t(\mathbf{x})$  corresponds to the radiance intensity captured at moment $t$ .

To learn accurate 3D representations from motion-blurred images, BAD-NeRF proposes linearly interpolating camera poses in $\textbf{SE}(3)$ between the initial pose  $\mathbf{T}_{\text{start}}$  at the exposure start and the final pose  $\mathbf{T}_{\text{end}}$  at the exposure end, generating sampled moments.:
\begin{equation}
	\mathbf{T}_t = \mathbf{T}_{\text{start}} \cdot \exp(\frac{t}{n}\cdot \log (\mathbf{T}_{\text{start}}^{-1} \cdot \mathbf{T}_{\text{end}}))
\end{equation}
At each sampled moment, NeRF renders a sharp image. The predicted motion-blurred image is obtained by averaging all rendered images across moments. The loss between this predicted image and the real captured one is computed to jointly optimize NeRF network, $\mathbf{T}_{\text{start}}$ , and $\mathbf{T}_{\text{end}}$ , minimizing the discrepancy between predicted and real motion blur.

\subsection{In-scattering Lightpath Model}
\begin{figure}[!h]
	\centering
	\includegraphics[width=0.45\textwidth]{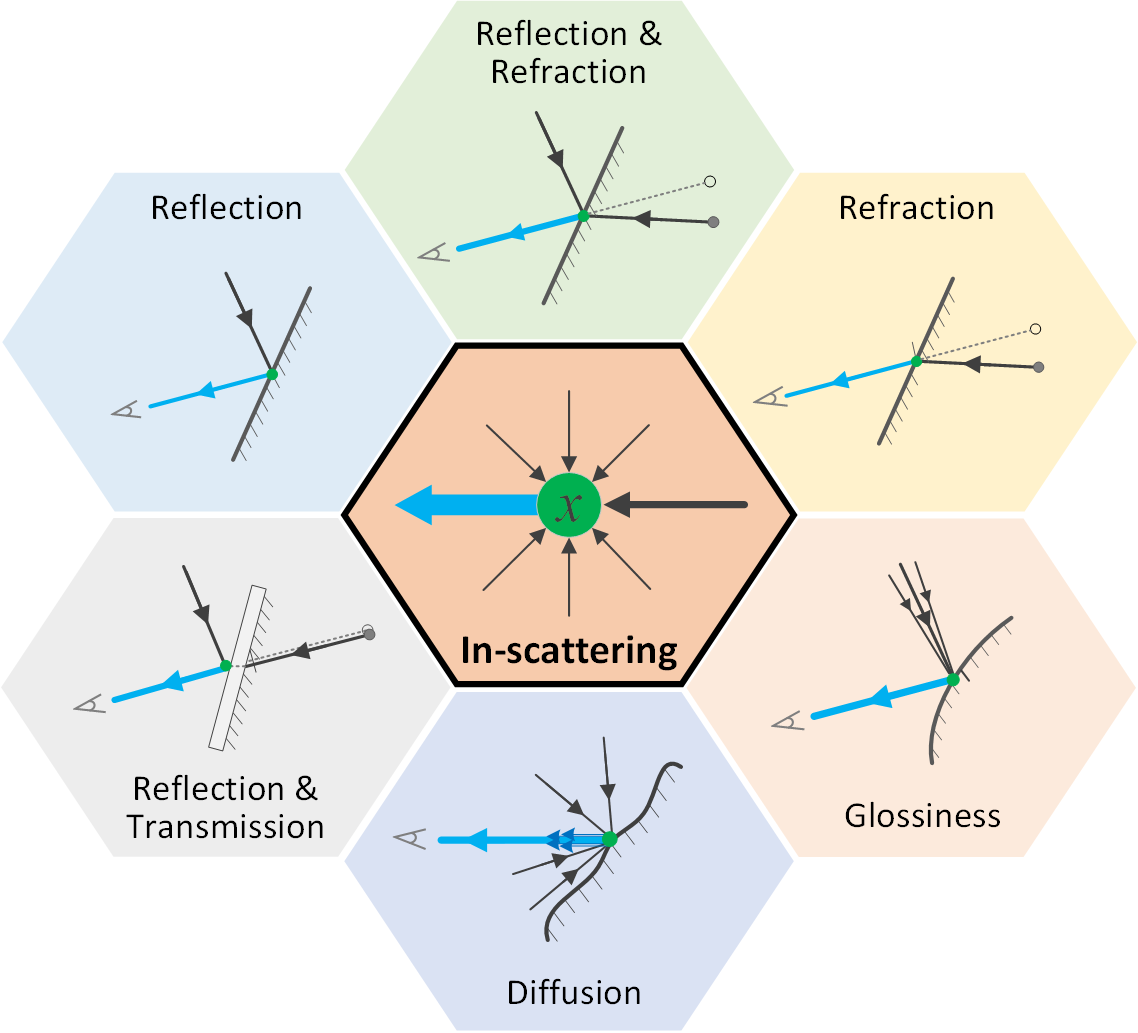}
	\caption{
		The in-scattering model of lightpaths.
	}
	\label{fig:fig_1}
\end{figure}

In-scattering Lightpath Model (ISLM) decomposes complex lightpaths based on physical principles and unifies them as light scattering. While multiple scattering is overly complex and unnecessary for our purposes, we focus on single in-scattering and simulate six common ways of light transportation using this model, see Figure \ref{fig:fig_1}.

\begin{enumerate}
	\item Specular Reflection. When light encounters a smooth surface (e.g., water, mirrors), its path reflects to form a polygonal path rather than a straight line before entering the camera.
	\item  Refraction. Light refracts when propagating between different media (e.g., observing underwater objects from the shore), altering its direction and causing a deviation in object’s position.
	\item  Reflection+Refraction. For instance, observing an underwater object alongside its reflected counterpart involves light paths combining reflection and refraction, where both paths converge to form the incident ray.
	\item  Reflection+Transmission. When viewing objects behind glass alongside their reflections (e.g., glass surfaces), the incident light combines reflected and transmitted paths. Transmission through thin glass can be approximated as straight-line propagation, while thicker glass necessitates refraction modeling.
	\item  Glossiness. A hybrid between specular and diffuse reflection, this phenomenon manifests as view-dependent highlights on surfaces like whiteboards or metals, where brightness varies with the observer’s angle.
	\item  Diffusion. Resulting from microscopic scattering, diffuse reflection aggregates light contributions from multiple directions, forming the incident ray observed at a pixel. This covers all non-specular surfaces encountered in daily observations.
\end{enumerate}
All six scenarios can be unified as in-scattering, where light scatters at a single point on a medium or surface, with multiple scattered rays converging into the camera’s incident path to form a pixel. 

\section{Appendix B}
\begin{figure*}[]
	\centering
	\includegraphics[width=1.0\textwidth]{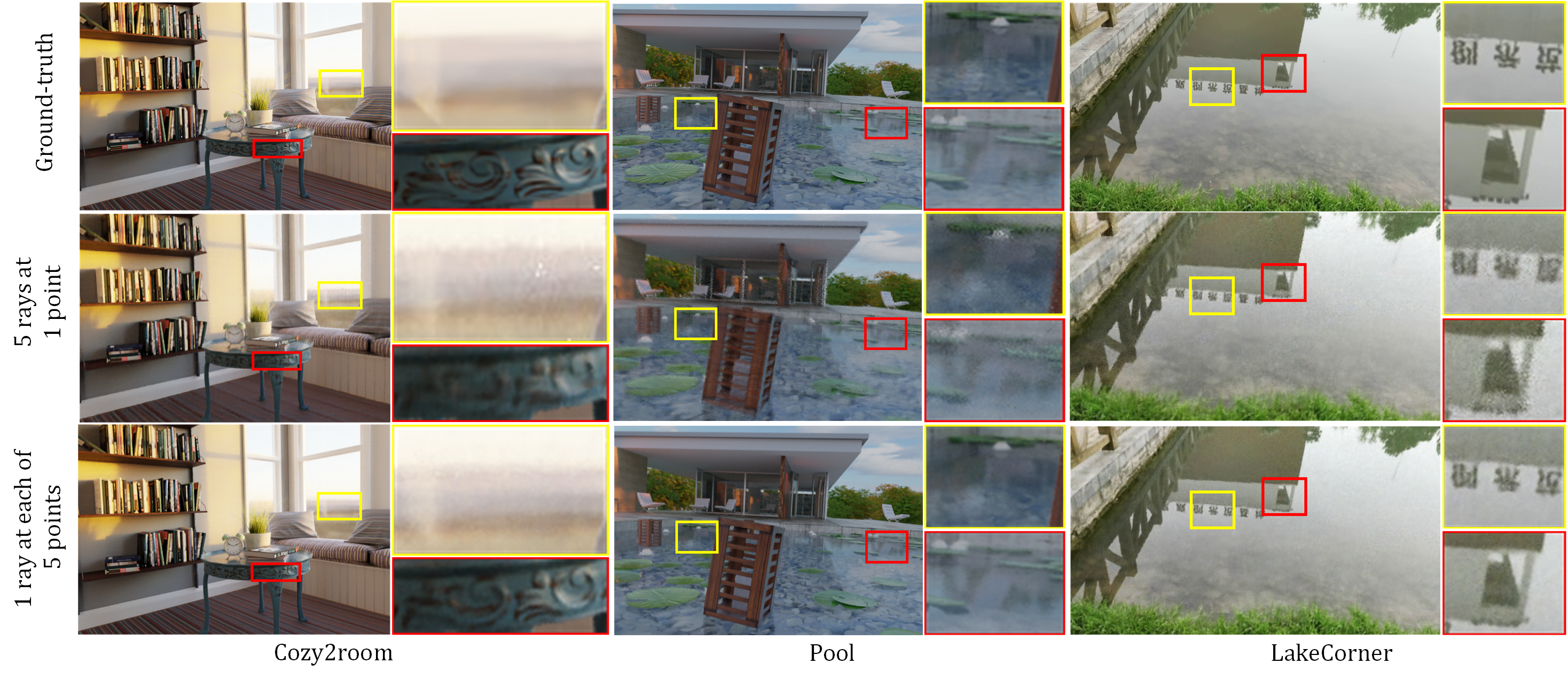}
	\caption{
		The intuitive comparisons of two scattering designs: 1) emitting five scattering rays from one primary sampling point, 2) emitting one scattering ray from each of five adjacent primary points. For example scenes of Deblur-dataset (Cozy2room and Pool) and self-captured dataset (LakeCorner), the second design evidently achieves better quality of synthesis images. The first design generates obvious noises while the second style captures more clearer details and handles light scatterings better (e.g. reflections on the water surface of Pool and LakeCorner, similar to ground-truth). 
	}
	\label{fig_app1}
\end{figure*}

Intuitive results of two scattering designs: 1) emitting five rays from a single primary sampling point, versus 2) emitting one scattering ray from each of five adjacent primary points. Emitting rays from five consecutive primary points outperforms the single-point multi-ray style, please see Figure \ref{fig_app1} for details.

\section{Appendix C}
\begin{figure*}[]
	\centering
	\includegraphics[width=1.0\textwidth]{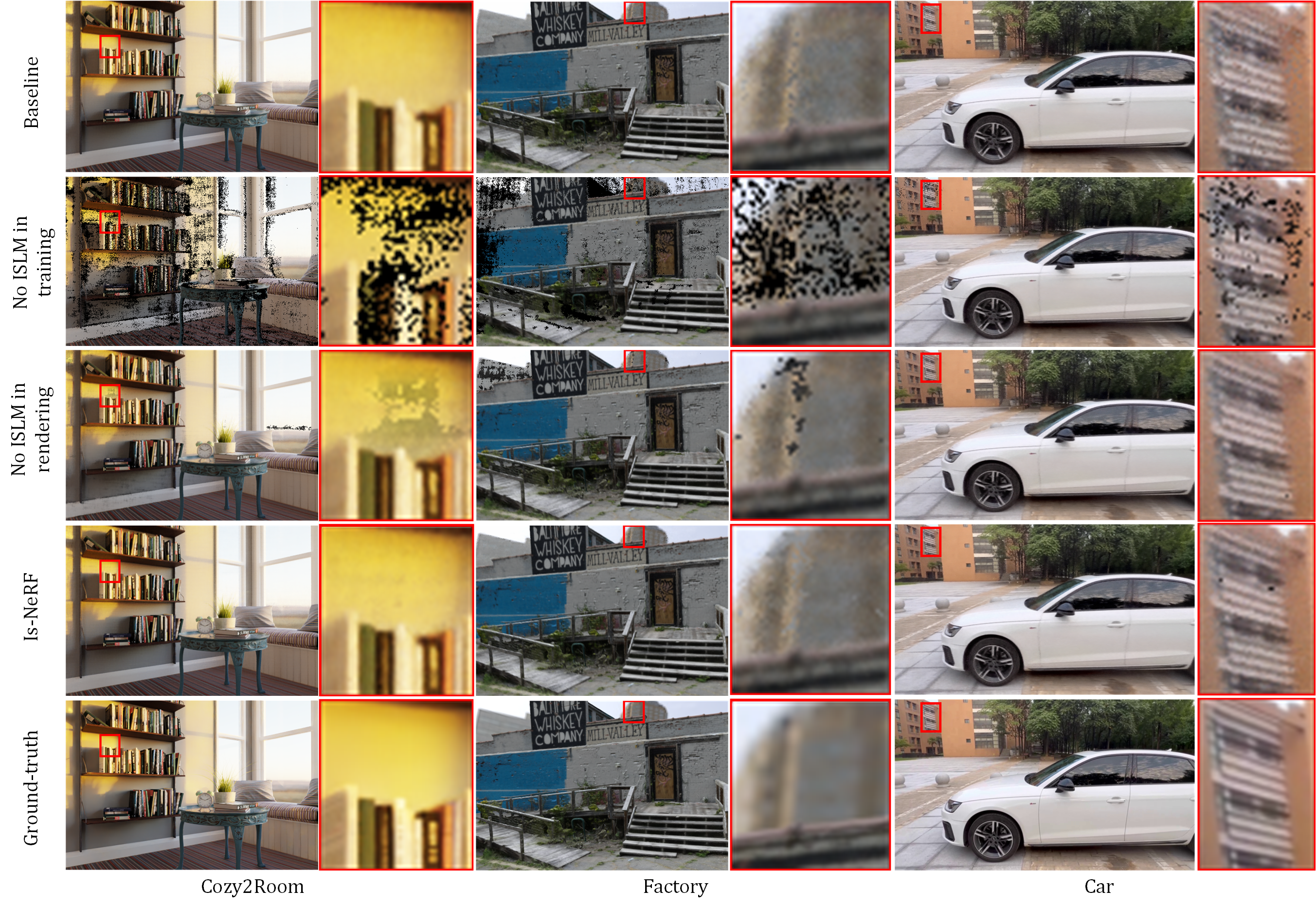}
	\caption{
		The intuitive results of ablation studies. By removing ISLM module from training, no available scatterings will be learned and then conducting scattering-aware volume rendering causes obvious data deficiency in the synthesis image of all three scenes (Cozy2room, Factory, and Car). While removing ISLM module from rendering leads to fewer flaws. By integrating ISLM in both training and rendering phases, our Is-NeRF significantly improves the quality of novel view images, superior to the baseline method. 
	}
	\label{fig_app2}
\end{figure*}

The intuitive results of ablation studies by removing ISLM module from training and rendering, respectively. Please see Figure \ref{fig_app2} for details.

\end{document}